\documentstyle[preprint,eqsecnum,aps,prd,epsf]{revtex}
\newcommand{\twocol}{\widetext}
\newcommand{\onecol}{\widetext}

\newcommand{\tG}{\tilde G}
\newcommand{\bx}{{\bf x}}
\newcommand{\bn}{{\bf n}}
\newcommand{\Viv}{{V^{\rm IV}}}
\newcommand{\VivR}{{V^{\rm IV}_{\rm ren}}}
\newcommand{\Vfv}{{V^{\rm FV}}}
\newcommand{\Deltax}{{\Delta x}}
\newcommand{\tr}{\mbox{\rm tr}\mbox{\bf 1}}
\newcommand{\m}{{m}}
\newcommand{\muM}{{\m_0}}
\newcommand{\zetan}{{\zeta_\bn}}
\newcommand{\sqrnn}{\sqrt{n_{1}^{2}+n_{2}^{2}}}
\newcommand{\sumnn}{n_{1}+n_{2}}
\newcommand{\mul}{{m_1}}
\newcommand{\LP}{{L_{\rm P}}}
\newcommand{\LA}{{L_{\rm A}}}

\makeatletter
\long\def\@makecaption#1#2{%
   \vskip 10\p@
   \setbox\@tempboxa\hbox{#1\ \ #2}%
   \ifdim \wd\@tempboxa >\hsize
   #1\ \ #2\par        %
      \else
   \hbox to\hsize{\hfil\box\@tempboxa\hfil}%
   \fi}
\def\fnum@figure{Fig. \thefigure}
\makeatother

\makeatletter
\def\figure{%
\let\@capwidth\columnwidth
\vskip1pc
\def\@captype{figure}%
\interlinepenalty10000 %
\@ifnextchar[{\@chuckoptarg}{}%
}%
\makeatother

\begin{document}

\draft
\titlepage
{\baselineskip0pt
\leftline{\large\baselineskip16pt\sl\vbox to0pt{\hbox{Department of Physics} 
               \hbox{Hiroshima University}\vss}}
\rightline{\large\baselineskip16pt\rm\vbox to20pt{\hbox{HUPD-9612} 
           \hbox{ICRR-Report 376-96-27}
\vss}}
}
\vspace*{3mm}
\begin{center}
{\large \bf Dynamical Symmetry Breaking in Flat Space
\\[2mm] with Non-trivial Topology}
\end{center}
\vspace{2mm}
\begin{center}
{Ken-ichi Ishikawa${}^{1}$,  Tomohiro Inagaki${}^{2}$, 
Kenji Fukazawa${}^{3}$, and Kazuhiro Yamamoto${}^{1}$}
\end{center}
\begin{center}
{\em ${}^{1}$Department of Physics,~Hiroshima University
         \\  Higashi-Hiroshima 739,~Japan}
\\
{\em ${}^{2}$Institute for Cosmic Ray Research,~University of Tokyo,~
         \\   Tanashi,~Tokyo 188,~Japan}
\\
{\em ${}^{3}$Department of Mechanical Engineering,~
  Kure National College of Technology, \\  Kure 737,~Japan}
\end{center}
\vspace{-8mm}
\begin{abstract} 
We consider a four-fermion theory as a simple model of dynamical 
symmetry breaking in flat space with non-trivial topology,
motivated from recent studies in similar considerations in curved space.
The phase structure is investigated, by developing a useful formalism
to evaluate the effective potential in arbitrary compactified flat 
space in 3- and 4-dimensional spacetime.  
The phase structure is significantly altered due to the finite
volume effect in the compactified space. 
Interestingly, the effect works in different way depending on 
the boundary condition of the fermion fields. 
The physical interpretation of the results
and its implication on the dynamical symmetry breaking
phenomenon in curved space are discussed.
\end{abstract}
\vspace{2mm}

\twocol
\section{Introduction}

{} Various phenomena associated with phase transitions at the early 
stage of the universe have been a subject of great interest in 
cosmology for two decades. These phase transitions of the
universe is motivated from the symmetry breaking phenomenon
in high energy particle physics. 
One of the decisive problems in high energy particle physics 
is how the model of unified theories can be tested.
It is expected that the primary symmetry of unified theories 
is broken down at the early universe to yield the theories 
with lower symmetries.
There is a possibility to test the model at the early stage
of the universe.

To investigate unified theories, much interest has been taken 
in clarifying the mechanism of the spontaneous symmetry breaking 
under the circumstance of the early universe.
The dynamics of the strong coupling gauge theory may break 
the symmetry of the unified theories without introducing 
an elementary scalar field.
This scenario is called as the dynamical symmetry breaking
\cite{NJL}.
It is considered
that the change of curvature or volume size of the universe
may cause the dynamical symmetry breaking as the evolution
of the universe proceeds.
The studies of such effects on the symmetry 
breaking may help understanding unified theories and evolution
of the early universe.

Many works have been done in this field. 
By using the weak curvature expansion it is found that the 
chiral symmetry is restored for a large positive curvature 
and that there is no symmetric phase in a spacetime with 
any negative curvature.\cite{IMO,ELOS,Inagaki}
In weakly curved spacetime it is pointed out that non-trivial 
topology for the fermion field may drastically change the 
phase structure of the four-fermion theory.\cite{ELO}
The higher derivative and gauged four-fermion theories
have also investigated in weakly curved spacetime.\cite{ENJL}
In some compact spaces, e.g., de Sitter space\cite{IMM,EOS} 
and Einstein universe\cite{IIM}, the effective potential 
is calculated without any approximation for the spacetime curvature.
It is observed to exhibit the symmetry restoration through the 
second order phase transition.
However, in such compact spaces, it is not clear whether the symmetry
restoration is caused by the curvature or finite size effect.
An example of other simple compact space with no curvature is 
the torus universe. 
Since the torus spacetime has only the finite size effect, 
then the investigation in this 
space will indicate which effects, curvature or finite size,
is essential to restore the symmetry.
We therefore investigate 
the dynamical symmetry breaking in compact flat space
with non-trivial topology. 

Let us briefly comment on the
cosmological motivations to consider the torus universe.
Several astrophysicists have discussed the possibility
of the torus universe \cite{Zeldovich,SS,FS}, and 
recently the topology of the universe is argued by 
using the observational data of the cosmic microwave background 
anisotropies \cite{torusCMB}, which was detected by COBE DMR \cite{smoot}.
Assuming that our universe is the three-torus, they
constrained the cell size of the torus.
According to the results, the size would be larger than the 
present horizon scale.
Thus we do not know the topology of our universe at present.
Nevertheless, there are some cosmological motivations to consider
compact flat space with non-trivial topology.
First, quantum cosmologist have argued that small volume 
universe have small action, and are more likely to be created 
\cite{Atkatz}. In fact, it seems difficult to create an infinite volume
universe in the context of quantum cosmology. 
Second, the torus universe, in contrast to the compact $S^3$ universe,
may have long lifetime because the curvature does not collapse 
the universe. 

In this paper we make a systematic study of the dynamical symmetry 
breaking in compact flat space with non-trivial topology,
assuming that the four-fermion theory is an effective
theory which stems from more fundamental theory at GUT era.
The effective potential is calculated from the Feynman
propagator which depends on the spacetime structure.
Evaluating the effective potential, we investigate the 
dynamical symmetry breaking induced by the effect of the 
spacetime structure.
The dynamical symmetry breaking in torus universe of space-time 
dimension, $D=3$, is investigated in Ref.\cite{Kimetal,KNSY,DYS}. 
Our strategy to evaluate the effective potential differs
from that in Ref.\cite{Kimetal,KNSY,DYS}. Our method starts
from the Feynman propagator in real space, then it
can be easily applied to compact flat spaces with arbitrary
topology for $D=2,3$ and $4$.

The paper is organized as follows.
In section 2, we show a brief review of four-fermion theory 
in curved space. We then extend the formalism 
to a useful form in order to investigate the effective potential 
in compact flat space with nontrivial topology.
In section 3, we apply the formalism to a 3-dimensional 
spacetime with nontrivial spatial sector.
4-dimensional case is investigated in section 4.
Section 5 is devoted to summary and discussions.
In appendix, we show the validity of our method by considering 
$D=2$ case. We prove that our method leads to
well known results previously obtained. 
We use the units $\hbar=1$ and $c=1$.

\section{Formalism}


In this section we first give a brief review of the
four-fermion theory
in curved space. We consider the system with the action\cite{GN}
\begin{equation}
     S\! =\!\! \int\!\!\! \sqrt{-g} d^{D}\! x\! \left[
     -\sum^{N}_{k=1}\bar{\psi}_{k}\gamma^{\alpha}\nabla_{\alpha}\psi_{k}
     +\frac{\lambda_0}{2N}
      \left(\sum^{N}_{k=1}\bar{\psi}_{k}\psi_{k}\right)^{2}
     \right] ,
\label{ac:gn}
\end{equation}
where index $k$ represents the flavors of the fermion field
$\psi$, $N$ is the number of fermion species, $g$ the determinant
of the metric tensor $g_{\mu\nu}$,
and $D$ the spacetime dimension.
For simplicity we neglect the flavor index below. 

The action (\ref{ac:gn}) is invariant under the discrete 
transformation
$\bar{\psi}\psi \longrightarrow -\bar{\psi}\psi$.
For $D=2,4$ this transformation is realized
by the  the discrete chiral transformation
$\psi \longrightarrow \gamma_{5}\psi$.
Thus we call this $Z^{2}$ symmetry the discrete chiral symmetry.
The discrete chiral symmetry prohibits the fermion mass term.
If the composite operator constructed from the fermion and
anti-fermion
develops the non-vanishing vacuum expectation value, 
$\langle\bar{\psi}\psi\rangle \neq 0$,
a fermion mass term appears in the four-fermion interaction
term and the chiral symmetry is broken down dynamically.

For practical calculations in four-fermion theory
it is more convenient to introduce auxiliary field $\sigma$
and start with the action
\begin{equation}
     S_{y} = \!\int\! \sqrt{-g}d^{D}x 
      \left[-\bar{\psi}\gamma^{\alpha}
      \nabla_{\alpha}\psi
     -\frac{N}{2\lambda_0}\sigma^{2}-\bar{\psi}\sigma\psi
      \right]\, .
\label{ac:yukawa}
\end{equation}
The action $S_{y}$ is equivalent to the action (\ref{ac:gn}).
If the non-vanishing vacuum expectation value is assigned to
the auxiliary field $\sigma$
there appears a mass term for the fermion field $\psi$
and the discrete chiral symmetry (the $Z_{2}$ symmetry)
is eventually broken.

We would like to find a ground state of the system described by 
the four-fermion theory.
For this purpose we evaluate an effective potential for the field 
$\sigma$.
The ground state is determined by observing the minimum of the
effective potential in the homogeneous and static background 
spacetime.

As is known, the effective potential 
in the leading order of the $1/N$ expansion is given by\cite{IMO}
\begin{equation}
  V(\sigma)={1\over 2\lambda_0}\sigma^2
  +{ {\rm Tr} \sqrt{-g} \int_0^\sigma ds S_{F}(x,x';s) 
    \over \int d^Dx \sqrt{-g}},
\label{def:epot}
\end{equation}
where 
\begin{equation}
  {\rm Tr}=\int\int d^Dx d^Dx' \delta^{D}(x-x') {\rm tr},
\end{equation}
and $S_{F}(x,x';s)$ is the Feynman propagator for free fermion 
with mass $s$, which satisfies
\begin{equation}
  (\gamma^\alpha\nabla_\alpha+s)S_{F}(x,x';s)={i\over\sqrt{-g}}
  \delta^D(x,x').
\end{equation}
It should be noted that the effective potential
(\ref{def:epot}) is normalized so that $V(0)=0$.

We introduce the Feynman propagator for the scalar 
field with mass $s$, 
\begin{equation}
  (\Box_x-s^2)G_{F}(x,x';s)=\frac{i}{\sqrt{-g}}\delta^D(x,x'),
\end{equation}
which has the relation,
\begin{equation}
  S_{F}(x,x';s)=(\gamma^\alpha\nabla_\alpha-s)G_{F}(x,x';s).
\end{equation}
Then, we write the effective potential as
\begin{equation}
  V(\sigma)={1\over 2\lambda_0}\sigma^2
  -\lim_{x'\rightarrow x }{\tr}\int_0^\sigma ds ~ s~G_{F}(x,x';s),
\label{expVb}
\end{equation}
in flat spacetime.
Here $\tr$ is the trace of an unit Dirac matrix.

Now we consider the Feynman propagator on compact flat space with 
nontrivial topology. We write the Feynman propagator in the 
$D$-dimensional Minkowski 
space as ${\tG}_{F}(x,x';s)=\tG(\xi)$, where
$\xi=(x-x')^2=(t-t')^2-({\bf x-x'})^2$ and the explicit expression
is given by Eq.(\ref{exptG}).
That is, ${\tG}_{F}(x,x';s)$ has the Lorentz invariance, 
and is a function of the variable $\xi$.
Then the Feynman propagator on the $(D-1)$-dimensional spatial 
torus whose size is ${\bf L}=(L_1,L_2,\cdots,L_{D-1})$ can be written 
\onecol
\begin{equation} 
  G_{F}(x,x';s)=
  \sum_{n_1=-\infty}^{\infty}\sum_{n_2=-\infty}^{\infty}
  \cdots \sum_{n_{D-1}=-\infty}^{\infty}\alpha(\bn) \tG(\xi_\bn)
  \equiv\sum_{\bn} \alpha(\bn) \tG(\xi_\bn),
\label{defGF}
\end{equation}
\twocol
where
\begin{equation} 
  \xi_\bn=(t-t')^2-\sum_{i=1}^{D-1}(x_i-x'_i+n_iL_i)^2,
\label{defxin}
\end{equation}
$\bn=(n_1,n_2, \cdots, n_{D-1})$, and
$\alpha(\bn)$ is a phase factor which is determined 
in accordance with the boundary condition of
quantum fields (see below).
Throughout this paper we use 
a convention $x=(t,\bx)=(t,x_1,x_2,\cdots,x_{D-1})$, 
$~x'=(t',\bx')=(t',x'_1,x'_2,\cdots,x'_{D-1})$, etc.
Note that the Green function constructed in this way has the invariance
under the replacement $x_i\rightarrow x_i +{ L_i}$, 
and satisfies the equation of motion.
\footnote{
We should note that our formalism will be easily extended to
finite temperature theory. 
The finite temperature Green function can be obtained
by summing the Euclidean Green function so that 
it has periodicity in the direction of time, 
in the same way as Eq.(\ref{defGF}).
}

The Feynman propagator in the $D$-dimensional Minkowski space is 
(see e.g.\cite{BD})
\onecol
\begin{equation}
  {\tG}_{F}(x,x';s)=\tG(\xi)={\pi\over (4\pi i)^{D/2}}
  \biggl({4s^2\over -\xi +i\epsilon}\biggr)^{(D-2)/4}
  H^{(2)}_{D/2-1}\Bigl([s^2(\xi-i\epsilon)]^{1/2}\Bigr),
\label{exptG}
\end{equation}
where $H_\nu^{(2)}(z)$ is the Hankel function of the second kind. 

Then the effective potential is obtained by substituting Eq.(\ref{exptG}) 
with (\ref{defxin}), and (\ref{defGF}) into Eq.(\ref{expVb}).
Performing the integration we get
\begin{equation}
  V={1\over 2\lambda_0}\sigma^2-\lim_{x'\rightarrow x }
  \tr\sum_{\bn}\alpha(\bn)
  {\pi\over(4\pi i)^{D/2}}
  \biggl({4\over -\xi_\bn+i\epsilon}\biggr)^{(D-2)/4} 
  {1\over (\xi_\bn-i\epsilon)^{1/2}}
  \biggl[ s^{D/2} H_{D/2}^{(2)}(s(\xi_\bn-i\epsilon)^{1/2})
  \biggr]_{0}^{\sigma}.
\end{equation}
The effective potential should take real values physically.
The imaginary part of the effective potential
must vanish after we take the limit $x \rightarrow x'$.
Thus we consider only real part of the effective potential which
is given by
\begin{equation}
  V={1\over 2\lambda_0}\sigma^2+\lim_{\bx'\rightarrow \bx }
  \tr\sum_{\bn}\alpha(\bn)
  {1\over(2\pi )^{D/2}}
  {1\over \Deltax_\bn^D}
  \biggl[ (\sigma\Deltax_\bn)^{D/2} K_{D/2}(\sigma\Deltax_\bn)
  -\lim_{z\rightarrow0} z^{D/2}K_{D/2}(z)\biggr]~,
\label{epb}
\end{equation}
\twocol
where we have defined
\begin{equation}
        \Deltax_\bn=\sqrt{\sum_{i=1}^{D-1}(x_i-x'_i+n_iL_i)^2}~,
\end{equation}
and $K_{\nu}(z)$ is the modified Bessel function.
In deriving the above equation, we have set that $t=t'$, and used the relation, $H_\nu^{(2)}(-iz)=(i2/\pi)e^{\nu\pi i} K_{\nu}(z)$
(see e.g. \cite{Mag}).

As is read from Eq.(\ref{epb}), only when $\bn=0$
in the summation diverges. 
We therefore separate the effective potential into two parts,
\begin{equation}
  V=\Viv+\Vfv,
\label{def:V:topo}
\end{equation}
where
\onecol
\begin{eqnarray}
  &&\Viv={1\over 2\lambda_0}\sigma^2+
  \lim_{\Deltax\rightarrow 0}
  \tr{1\over(2\pi )^{D/2}}
  {1\over \Deltax^D}
  \biggl[ (\sigma\Deltax )^{D/2} K_{D/2}(\sigma\Deltax)
  -2^{D/2-1}\Gamma(D/2)\biggr],
\label{exViv}
\\
    &&\Vfv=\lim_{\bx'\rightarrow \bx }
  \tr\sum_{\bn (\ne 0)}\alpha(\bn)
  {1\over(2\pi )^{D/2}}
  {1\over \Deltax_\bn^D}
  \biggl[ (\sigma\Deltax_\bn)^{D/2} K_{D/2}(\sigma\Deltax_\bn)
  -2^{D/2-1}\Gamma(D/2)\biggr]~.
\label{exVfv}
\end{eqnarray}
\twocol
Here we have set $\alpha(\bn=0)=1$,
and used that $\lim_{z\rightarrow0} z^{D/2}K_{D/2}(z)=2^{D/2-1}\Gamma(D/2)$.

In general, we need to regularize the divergence of $\Viv$
by performing the renormalization procedure.
It is well known that the divergence can be
removed by the renormalization of the 
coupling constant for $D<4$. 
Employing the renormalization condition,
\begin{equation}
  {\partial^2 \Viv\over \partial \sigma^2}~\bigg\vert_{\sigma=\mu} = 
  {\mu ^{D-2}\over \lambda_r},
\end{equation}
we find that Eq.(\ref{exViv}) reads
\onecol
\begin{eqnarray}
  &&\VivR={1\over{2\lambda_r}}\sigma^2 \mu^{D-2}
\nonumber
\\
  &&\hspace{0.5cm}
  +\lim_{\Deltax\rightarrow0} {\tr\over(2\pi)^{D/2}}
  {1\over \Deltax^D}
  \biggl[{1\over2}(\sigma\Deltax)^2(\mu\Deltax)^{D/2-1}
  \Bigl(K_{D/2-1}(\mu\Deltax)-\mu\Deltax K_{D/2-2}(\mu\Deltax)\Bigr)
\nonumber
\\
  &&\hspace{4.6cm}+(\sigma\Deltax)^{D/2}K_{D/2}(\sigma\Deltax)
  -2^{D/2-1}\Gamma(D/2)
  \biggr].
\label{Vivren}
\end{eqnarray}
\twocol
As we shall see in the next sections, $\VivR$ reduces to
the well known form of the effective potential in the Minkowski 
spacetime. Therefore the effect of the nontrivial configuration 
of space on the effective potential is described by $\Vfv$.

Finally in this section, we explain the phase factor $\alpha(\bn)$.
As is pointed out in Ref.\cite{ISHAM,Dowker,Avis}
there is no theoretical constraint which boundary
condition one should take for quantum fields in compact flat spaces.
It is possible to consider the fields with the various boundary 
condition in the compact spaces with non-trivial topology.
Thus we consider the fermion fields with periodic and
antiperiodic boundary conditions, and
study whether the finite size effect can be changed by the 
boundary condition. For this purpose, it is convenient to 
introduce the phase factor $\alpha(\bn)$ by
\begin{equation}
  \alpha(\bn)=\alpha_1\alpha_2\cdots \alpha_{D-1},
\end{equation}
where $\alpha_i=(-1)^{n_i}$ for antiperiodic boundary condition 
in the direction of $x_i$, and $\alpha_i=1$ 
for periodic boundary condition.
In the following sections we investigate the behavior of the effective
potential at $D=3$ and $D=4$ with the various boundary conditions. 

\section{Application in $D=3$}

In this section
we apply the method explained in the previous section to 
the case, $D=3$.
In the three dimensional torus spacetime, $R\otimes S^1\otimes S^1$,
it is possible to consider the three kinds of independent boundary 
conditions.
To see the effect of the compact space and the boundary condition 
of the field on the phase structure of the four-fermion theory,
we evaluate the effective potential and the gap equation 
and show the phase structure for every three kinds of boundary conditions.
As is mentioned before, the same problem has been investigated
in Ref.\cite{Kimetal} for the three dimensional flat compact space
with nontrivial topology.
We can compare our results with theirs.

The three dimensional case is instructive because the 
modified Bessel functions reduce to elementary functions,
\begin{eqnarray}
  &&K_{3/2}(z)=\sqrt{\pi\over 2z} \biggl(1+{1\over z}\biggr)e^{-z},
\label{eqn:bes1}\\
  &&K_{1/2}(z)=K_{-1/2}(z)=\sqrt{\pi\over 2z}e^{-z}.
\label{eqn:bes2}
\end{eqnarray}
Substituting the relations (\ref{eqn:bes1}) and (\ref{eqn:bes2})
to Eqs.(\ref{Vivren}) and (\ref{exVfv}),
the effective potential $V$ in three dimension
becomes
\begin{equation}
 V=\VivR + \Vfv,
\label{eqn:Effpt3D}
\end{equation}
\begin{equation}
  {\VivR\over \mu^3}=
\frac{1}{2}\left(\frac{1}{\lambda_{r}}-\frac{\tr}{2\pi}\right)
  \left(\frac{\sigma}{\mu}\right)^{2}
  + \frac{\tr}{12\pi}\left(\frac{\sigma}{\mu}\right)^{3},
\label{Vivc3}
\end{equation}
\begin{equation}
\frac{\Vfv}{\mu^{3}}=\frac{\tr}{4\pi}\sum_{\bn\neq0}
\frac{\alpha(\bn)}{(\mu\zetan)^{3}}
\left(\sigma\zetan e^{-\sigma\zetan}+e^{-\sigma\zetan}-1\right),
\label{eqn:Vfv3d}
\end{equation}
where
\begin{equation}
  \zetan \equiv\lim_{\bx'\rightarrow\bx}\Deltax_\bn
  =\sqrt{(n_1L_1)^2+(n_2L_2)^2}.
\label{defzetan}
\end{equation}
Here we note that Eq.(\ref{eqn:Vfv3d}) disappears in the
Minkowski limit $(L_{1},L_{2})\rightarrow (\infty,\infty)$
and the effective potential (\ref{eqn:Effpt3D}) 
reduces to Eq.(\ref{Vivc3}) which is equal to the effective
potential in the Minkowski space-time.

The gap equation,
 $\partial V /\partial\sigma\vert_{\sigma=m}=0$,
 which determines the dynamical fermion mass $m$
 in the compact flat space with nontrivial topology
 reduces to
\begin{equation}
\frac{4\pi}{\lambda_{r}\tr}-2+\frac{m}{\mu}-
\sum_{\bn\neq0}\alpha(\bn)\frac{e^{-m\zetan}}{\mu\zetan}=0.
\label{eqn:Gap3D1}
\end{equation}
The effective potential in the Minkowski spacetime (\ref{Vivc3})
has a broken phase in which the discrete 
chiral symmetry is broken down, when the coupling
constant is larger than the critical value $\lambda_{cr}=2\pi/\tr$. 
For convenience, we introduce the dynamical fermion mass $\muM$ in 
the Minkowski space-time given by
$\partial\VivR/ \partial\sigma\vert_{\sigma=\muM}=0$. 
The dynamical fermion mass $m_{0}$ in the Minkowski 
space-time has a relationship with
the coupling constant as
\begin{equation}
\frac{m_{0}}{\mu}=-\frac{4\pi}{\lambda_{r}\tr}+2.
\label{eqn:m0vscpl}
\end{equation}
When the system is in the broken phase at the limit of Minkowski space
$(L_{1},L_{2})\rightarrow(\infty,\infty)$,
substituting Eq.(\ref{eqn:m0vscpl})
to Eq.(\ref{eqn:Gap3D1}),
we obtain the gap equation,
\begin{equation}
\frac{m}{m_{0}}-1-
\sum_{\bn\neq0}\alpha(\bn)\frac{e^{-m\zetan}}{m_{0}\zetan}=0.
\label{eqn:Gap3D2}
\end{equation}
It should be noted that the solution $m$ of the gap equation 
coincides with the dynamical fermion mass $m=m_{0}$ at the 
limit $(L_{1},L_{2})\rightarrow(\infty,\infty)$.

We expect that the phase transition will occur for
a sufficiently small $L_{i}$.
It is convenient to introduce a new variable $k_{i}$ instead
of $L_{i}$ to investigate the gap equation (\ref{eqn:Gap3D2}) for
small $L_{1}$ and $L_{2}$.
\begin{equation}
  k_{i} \equiv \left(\frac{2\pi}{L_{i}}\right)^{2}.
\end{equation}
To investigate the phase structure of the four-fermion theory
in three dimensional flat compact space, 
we calculate the effective potential
(\ref{eqn:Effpt3D}) and the gap equation (\ref{eqn:Gap3D2})
numerically with varying the variable $k_{i}$ for the various
boundary conditions below.

\subsection{Antiperiodic-antiperiodic boundary condition}
First we take the antiperiodic boundary condition for both
compactified directions and call this case AA-model.
The phase factor is chosen as
$\alpha(\bn) = (-1)^{n_{1}}(-1)^{n_{2}}$ in this case. 
In Fig.\ref{fig:Gap3dAA} the behavior of the gap equation
(\ref{eqn:Gap3D2}) is shown for the AA-model. 
As is seen in Fig.\ref{fig:Gap3dAA}, we find that 
the symmetry restoration occurs as $L_{1}$ and (or) $L_{2}$
become smaller and that the phase transition is second-order.
In the case of $L_{1}=L_{2}=L$,
the critical value of $L$ where the phase transition takes place is 
$L_{cr}\approx 1.62/m_{0} $ ($k_{cr}/{m_{0}}^{2} \approx 15.1$).
In Fig.\ref{fig:EffptAA} the behavior of the effective
potential (\ref{eqn:Effpt3D}) for the AA-model
is plotted as a function of $\sigma/\mu$ 
for the case of $\lambda_{r}>\lambda_{cr}$ (we take 
$\lambda_{r}=2 \lambda_{cr}$ as a typical case).
In plotting Figs.\ref{fig:Gap3dAA} and \ref{fig:EffptAA},
we numerically summed $\bn$ in
Eqs.(\ref{eqn:Vfv3d}) and (\ref{eqn:Gap3D2}).
In Fig.\ref{fig:PhaseAALxLy} we show the phase
diagram for the AA-model in $(L_{1},L_{2})$ plane. 
Here we note that in the limit 
$L_{1} \rightarrow \infty$ (or $L_{2} \rightarrow \infty$), the
space-time topology $R\otimes S^1 \otimes S^1$,
considered here, should be understood as $R^{2}\otimes S^1$. 
In this limit the field theory should have the same structure 
as the finite temperature field theory for $D=3$.
In fact, the critical value of $L_{1}$ (or $L_{2}$) is 
equal to the critical temperature $\beta_{cr}= 2\ln 2 
\approx 1.39$ as is expected, which is shown by dashed line
in the figure.
These results consistent with those of Ref.\cite{Kimetal}
for the model with the A-A boundary condition.

%
%
%

\subsection{Periodic-antiperiodic boundary condition}
We consider the case where the periodic
and the antiperiodic boundary conditions are adopted
in $x_{1}$ and  $x_{2}$ directions, respectively.
We call this case PA-model, where
the phase is taken as $\alpha(\bn)=(-1)^{n_{2}}$. 
In Fig.\ref{fig:Gap3dPA} we show the behavior of the gap 
equation (\ref{eqn:Gap3D2}) for the PA-model.
From Fig.\ref{fig:Gap3dPA}, we find that the symmetry restoration 
occurs when $L_{2}$ becomes smaller with $L_{1}$ fixed
and that the symmetry restoration dose not occur
when $L_{1}$ becomes smaller with $L_{2}$ fixed.
We also find that the phase transition is second-order, if occur.
Especially in the case of $L_{1}=L_{2}=L$, the symmetry restoration 
occur, and the critical value $L_{cr}$ where the symmetry is restored 
is $m_{0}L_{cr} \approx 1.14$ ($k_{cr}/m_{0}^{2} \approx 30.3$).
In Figs.\ref{fig:EffptPA}, \ref{fig:EffptPAKx10KyetcAP} and
\ref{fig:EffptPAKxetcKy10AP}
typical behaviors of the effective potential
for the PA-model are shown as a function of $\sigma/\mu$ 
for the case of $\lambda_{r}>\lambda_{cr}$ 
(we take $\lambda_{r}=2 \lambda_{cr}$). 
Fig.\ref{fig:EffptPA} is the case of $L_{1}=L_{2}$.
Fig.\ref{fig:EffptPAKx10KyetcAP} is the case with $L_{1}$ fixed,
and  Fig.\ref{fig:EffptPAKxetcKy10AP} is same but with $L_{2}$ fixed.
In Fig.\ref{fig:EffptPAKx10KyetcAP}, where $L_1$ (the size
associated with the periodic boundary condition) is fixed,
we can see that the chiral symmetry is restored as $L_2$ (the size associated 
with the anti-priodic boundary condition) becomes smaller.
While, in Fig.\ref{fig:EffptPAKxetcKy10AP} where $L_2$ is
fixed, the fermion mass becomes larger as $L_1$ becomes smaller.
The phase diagram for the 
PA-model is shown in Fig.\ref{fig:PhasePALxLy} in $(L_{1},L_{2})$ plane.
These results is consistent with those of Ref.\cite{Kimetal}.

\subsection{Periodic-periodic boundary condition}
Finally we take the periodic boundary condition for both compactified
directions and call this case PP-model.
The phase factor is chosen as $\alpha(\bn)=1$ in this case. 
In Fig.\ref{fig:Gap3dPP}, we show the behavior of the gap equation 
(\ref{eqn:Gap3D2}) for the PP-model. From this figure,
we find that the symmetry restoration does not occur as $L_{1}$
and (or) $L_{2}$ becomes smaller. In Fig.\ref{fig:EffptPP}, we show
the effective potential (\ref{eqn:Effpt3D}) in the case of 
$L_{1}=L_{2}=L$ and $\lambda_{r}>\lambda_{cr}$ 
(we take $\lambda_{r}=2\lambda_{cr}$) for the PP-model. 

Especially, in the case of $L_{1}=L_{2}=L$, 
we can analytically prove that the symmetry restoration doesn't 
occur irrespective of the coupling constant $\lambda_{r}$.
To see this,  we investigate the differential coefficient 
of the effective potential (\ref{eqn:Effpt3D}) of the PP-model
at $\sigma \rightarrow 0+$. The differential coefficient of the
effective potential (\ref{eqn:Effpt3D}) is
\begin{equation}
\frac{1}{\mu^{2}}\frac{dV}{d\sigma}
 =
\frac{\sigma}{\mu}\left(
\frac{1}{\lambda_{r}}-\frac{\tr}{2\pi}+
\frac{\tr}{4\pi}\frac{\sigma}{\mu}
-\frac{\tr}{4\pi}\sum_{\bn\neq0}
\frac{e^{-\sigma\zetan}}{\mu\zetan} \right).
\label{eqn:coeff}
\end{equation}
Taking the limit $\sigma \rightarrow 0+$, Eq.(\ref{eqn:coeff}) 
reduces to
\begin{equation}
\left.
\frac{1}{\mu^{2}}\frac{dV}{d\sigma}
\right|_{\sigma\rightarrow 0+}
=
-\frac{\tr \sigma}{\mu\pi}\left.
\sum_{n_{1}=1}^{\infty}\sum_{n_{2}=1}^{\infty}
\frac{e^{-\sigma L\sqrnn}}{\mu L\sqrnn}
\right|_{\sigma\rightarrow 0+}.
\label{eqn:prof1}
\end{equation}
Using an inequality,
\begin{equation}
\frac{\sumnn}{\sqrt{2}} \leq \sqrnn < \sumnn,
\mbox{for $n_{1}\geq 1$ and $n_{2}\geq 1$},
\end{equation}
we get the inequality
\begin{equation}
\frac{e^{-\sigma L(\sumnn)}}{\sumnn} <
\frac{e^{-\sigma L\sqrnn}}{\sqrnn} \leq
\frac{\sqrt{2}e^{-{\sigma L}(\sumnn)/{\sqrt{2}}}}{\sumnn}.
\label{eqn:ineq1}
\end{equation}
Summing up each term in Eq.(\ref{eqn:ineq1}) 
with respect to $n_{1}$ and $n_{2}$, we find,
\onecol
\begin{equation}
\frac{1}{e^{\sigma L}-1}+\log\left(1-e^{-\sigma L}\right) <
\sum_{n_{1}=1}^{\infty}\sum_{n_{2}=1}^{\infty}
\frac{e^{-\sigma L\sqrnn}}{\sqrnn} \leq
\frac{\sqrt{2}}{e^{{\sigma L}/{\sqrt{2}}}-1}
+\sqrt{2}
\log\left(1-e^{-{\sigma L}/{\sqrt{2}}}\right).
\label{eqn:ineq2}
\end{equation}
\twocol
According to Eq.(\ref{eqn:ineq2}), we obtain the inequality
\onecol
\begin{equation}
\left.\frac{1}{\mu^{2}}\frac{dV}{d\sigma}
\right|_{\sigma\rightarrow 0+} 
<
\left. -\frac{\tr\sigma}{\pi\mu^{2}L}\left(
\frac{1}{e^{\sigma L}-1}+\log\left(1-e^{-\sigma L}\right)
\right)\right|_{\sigma\rightarrow 0+}
= -\frac{\tr}{\pi\mu^{2}L^{2}} < 0.
\end{equation}
\twocol
We find that the differential coefficient of the effective potential
has a negative value at $\sigma \rightarrow 0+$
irrespective of  the coupling constant $\lambda_{r}$
in the case of $L_{1}=L_{2}=L$ for the PP-model.
Thus, we have shown that only a broken phase could exist and 
the symmetry  restoration does not occur at all.
Though this proof is limited to the special case $L_{1}=L_{2}=L$,
this result is expected to hold in other cases $L_{1} \neq \L_{2}$.
The results in this subsection are different from those of Ref.\cite{Kimetal}
for PP-model.

%

Summarizing this section, we examined the phase structure
of the four-fermion theory in compactified space of 
the three dimension by evaluating the effective
potential.
The phase structure is altered due to the compactified
space. Our results are consistent with those
of Ref.\cite{Kimetal} except for the periodic-periodic
boundary condition.
In the case of PP-model, our results indicate that 
only a broken phase could exist and the symmetry restoration 
dose not occur.

The behavior of the dynamical fermion mass $m$ is quite 
different according as the imposed boundary condition.
Making the length size of the compactified direction small, 
the dynamical fermion mass becomes large when adopting the 
periodic boundary condition. However it becomes small 
when adopting the antiperiodic boundary condition.
In concrete, for the AA-model where the antiperiodic boundary 
condition is imposed in the two directions,
the dynamical fermion mass disappears, and the symmetry is
restored when the size of the compactified space becomes small.
The order of the symmetry restoration is  second.
In the PP-model, where the periodic boundary
condition is imposed, the symmetry is not restored 
as is seen before.
In the PA-model, where one direction is periodic 
boundary condition and the other is antiperiodic one,
two effects compete with each other.
In the special case $L_{1}=L_{2}=L$ 
the effect of antiperiodic boundary condition
triumphs over that of periodic boundary condition
to restore the symmetry at small $L$.
The order of this symmetry restoration is  second.


\section{Application in $D=4$}

In this section we consider the case of  $D=4$, i.e.,
the spacetime which has 3-dimensional spatial sector 
with nontrivial topology. Let us start evaluating $\VivR$.
The special situation in $D=4$ case is that the 
renormalization can not make finite the effective potential
in our theory.
Therefore it is not necessary to consider the renormalization 
for $D=4$. Nevertheless, we introduce a "renormalized" coupling 
constant defined by Eq.(\ref{Vivren}) for convenience in the same way as 
the case in $D=3$ .

Then we must regularize $\VivR$ by some method,
e.g., by introducing a cut-off parameter.
Here we examine two methods to regularize $\VivR$.
The first one is to keep $\Deltax$ finite and to set $D=4$.
The straightforward calculations lead to
\onecol
\begin{eqnarray}
  {\VivR\over \mu^4}={1\over 2\lambda_{r}}\biggl({\sigma\over\mu}\biggr)^2
  + && {\tr\over4(4\pi)^2}
  \biggl[
  \biggl({\sigma\over\mu}\biggr)^2
  \biggl(-2+12\Bigl(\gamma-\ln{2\over\mu\Deltax}\Bigr)\biggr)
\nonumber \\
 && +\biggl({\sigma\over\mu}\biggr)^4
  \biggl({3\over2}-2\Bigl(\gamma-\ln{2\over\mu\Deltax}+\ln{\sigma\over\mu}
  \Bigr)\biggr)
  \biggr] +O(\Deltax),
\label{Vivc}
\end{eqnarray}
from Eq.(\ref{Vivren}).
On the other hand one can adopt the dimensional regularization 
as the second method, in which we
set $D=4-2\epsilon$. By expanding the right hand side of 
Eq.(\ref{Vivren}) in terms of $1/\epsilon$, we find
\begin{eqnarray}
  {\VivR\over \mu^D}={1\over 2\lambda_r}\biggl({\sigma\over\mu}\biggr)^2
  + &&{\tr\over4(4\pi)^2}
  \biggl[
  \biggl({\sigma\over\mu}\biggr)^2
  \biggl(-2+6\Bigl(\gamma-\ln 4\pi\Bigr) -{6\over\epsilon}\biggr)
\nonumber  
\\
&&+\biggl({\sigma\over\mu}\biggr)^4
  \biggl({3\over2}-\Bigl(\gamma-\ln4\pi + 2\ln {\sigma\over\mu}\Bigr)
  +{1\over\epsilon}\biggr)
  \biggr]+O(\epsilon).
\label{Vivcc}
\end{eqnarray}
\twocol
The above two methods are related  by
\begin{equation}
  {1\over\epsilon}=-\gamma+\ln\biggl({1\over (\mu\Deltax)^2\pi}\biggr).
\end{equation}
According to the Ref.\cite{Inagaki,IMO}, we find the momentum
cut-off parameter $\Lambda$, which is introduced in their papers,
is related by
\begin{equation}
  \ln{\Lambda^2\over \mu^2}=\ln\biggl({2\over \mu\Deltax}\biggr)^2
  +1-2\gamma.
\end{equation}

This effective potential has a broken phase, when the coupling
constant is larger than the critical value $\lambda_{\rm cr}$,
which is given by
\begin{equation}
  {1\over\lambda_{\rm cr}}=
  {\tr\over(4\pi)^2}\biggl(3\ln{\Lambda^2\over\mu^2}-2\biggr).
\end{equation}
For convenience, 
we introduce the dynamical fermion mass $\muM$ in the Minkowski space by
\begin{equation}
  {1\over\lambda_{\rm r}}-{1\over\lambda_{\rm cr}}
  +{\tr\over(4\pi)^2}\biggl(\ln{\Lambda^2\over\muM^2}\biggr)
  {\muM^2\over\mu^2}=0,
\end{equation}
as in the case in $D=3$.
In terms of $\muM$, Eq.(\ref{Vivc}) or Eq.(\ref{Vivcc})
 can be written as
\begin{equation}
 {\VivR\over\muM^4}={\tr\over4(4\pi)^2}\biggl[
  -2\biggl(\ln{\Lambda^2\over\muM^2}\biggr)~{{\sigma^2\over\muM^2}}
 + \biggl(\ln {\Lambda^2\over\muM^2}
 +{1\over2}-\ln {\sigma^2\over\muM^2}\biggr)
  {\sigma^4\over\muM^4}
  \biggr],
\end{equation}
where we used the momentum cut-off parameter.

On the other hand, Eq.(\ref{exVfv}), which represents the effect due to the 
compactified space,  reduces to
\begin{equation}
  {\Vfv\over\muM^4}=
  {\tr\over(2\pi)^2}\sum_{\bn\neq0} {\alpha(\bn)\over (\muM\zetan)^4}
  \biggl[ (\sigma\zetan)^2 K_2(\sigma\zetan)-2\biggr],
\end{equation}
where $\zetan=\sqrt{(n_1L_1)^2+(n_2L_2)^2+(n_3L_3)^2}$.

Then we get the gap equation 
\begin{equation}
-\ln{\Lambda^2\over\muM^2}
  +{\m^2\over\muM^2}\biggl(\ln{\Lambda^2\over\muM^2}-
  \ln{\m^2\over\muM^2}\biggr)
  -4\sum_{\bn\neq0}{\alpha(\bn)\over(\muM\zetan)}
  {\m\over\muM}K_{1}(\m\zetan)=0~.
\end{equation}
We can easily solve the above equation numerically.
The advantage of our method is that the equation is 
given by a simple sum of the modified Bessel function which 
damps exponentially at large $\bn$.

In $D=4$ case, we have many varieties of models
according to the varieties of the size of torus and
the boundary condition of fermion fields in the three different 
directions of torus.
For convenience, we separate this section into the following 
three subsections.

\subsection{Antiperiodic boundary conditions}

Let us first consider the models associated with the antiperiodic
boundary condition for fermion fields, where the phase parameter 
is given by $\alpha(\bn)=(-1)^{n_1}(-1)^{n_2}(-1)^{n_3}$.
In Fig.\ref{fig:figgapA} we show the behavior of solutions of the gap equation with 
the antiperiodic boundary condition on the three typical spaces with 
non-trivial topology. That is,
one is the torus of three equal sides ($L_1=L_2=L_3=L$), which we call
this case AAA-model,
second is the torus but with a infinite side, i.e., ($L_1=L_2=L$,
$L_3=\infty$),
AAI-model,
third is the space with only one side compactified ($L_1=L$,
$L_2=L_3=\infty$), which we call AII-model.
The three lines in the figure show 
the solution of the gap equations on the three spaces as the function
of $(2\pi/L m_0)^2$. 
For a direction $x_i$ with a infinite side, the sum of $n_i$ 
becomes ineffective.
Here we have taken the cut-off parameter $\Lambda/\muM=10$.\footnote{
In the below, the cut-off parameter  $\Lambda/\muM=10$  is adopted
tacitly, as long as we do not note especially.}

In Fig.\ref{fig:figgapA} we have considered the models which have 
broken phase for large $L$.
The figure shows that the symmetric phase appears as $L$ become
smaller beyond some critical values. Thus as is expected 
from the results of $D=3$, the effect of compactifying the space
affects the phase structure, and  the effect with the 
antiperiodic boundary condition for fermion fields always 
tends to restore the symmetry.

We show the critical values $L_{\rm cr}$, i.e.,
the value of $L$ when the symmetry is restored, 
as a function  of cut-off parameter in Fig.\ref{fig:figcrtA}. We can read
from the figure that smaller value of $L$ is needed in order
to restore the symmetry of the AII-model compared with AAA-model . 

%
%

\subsection{Periodic boundary conditions}

Next we consider the fields with the periodic boundary condition,
where the phase parameter is taken as $\alpha(\bn)=(+1)^{n_1}
(+1)^{n_2}(+1)^{n_3}$. In the same way as the above, consider
three typical kinds of spaces,
($L_1=L_2=L_3=L$), ($L_1=L_2=L$, $L_3=\infty$), and
($L_1=L$, $L_2=L_3=\infty$), with adopting the periodic boundary 
condition for the compactified directions.
We call each of them PPP-, PPI-, PII-model, respectively.
The solutions of gap equation are shown in Fig.\ref{fig:figgapP}.
In contrast to the results of antiperiodic boundary condition,
the fermion mass of broken phase becomes larger as $L$ becomes
smaller. The effect becomes more significant  as the scale of 
compactification $L$ becomes smaller. 
This nature is same as the results in $D=3$.

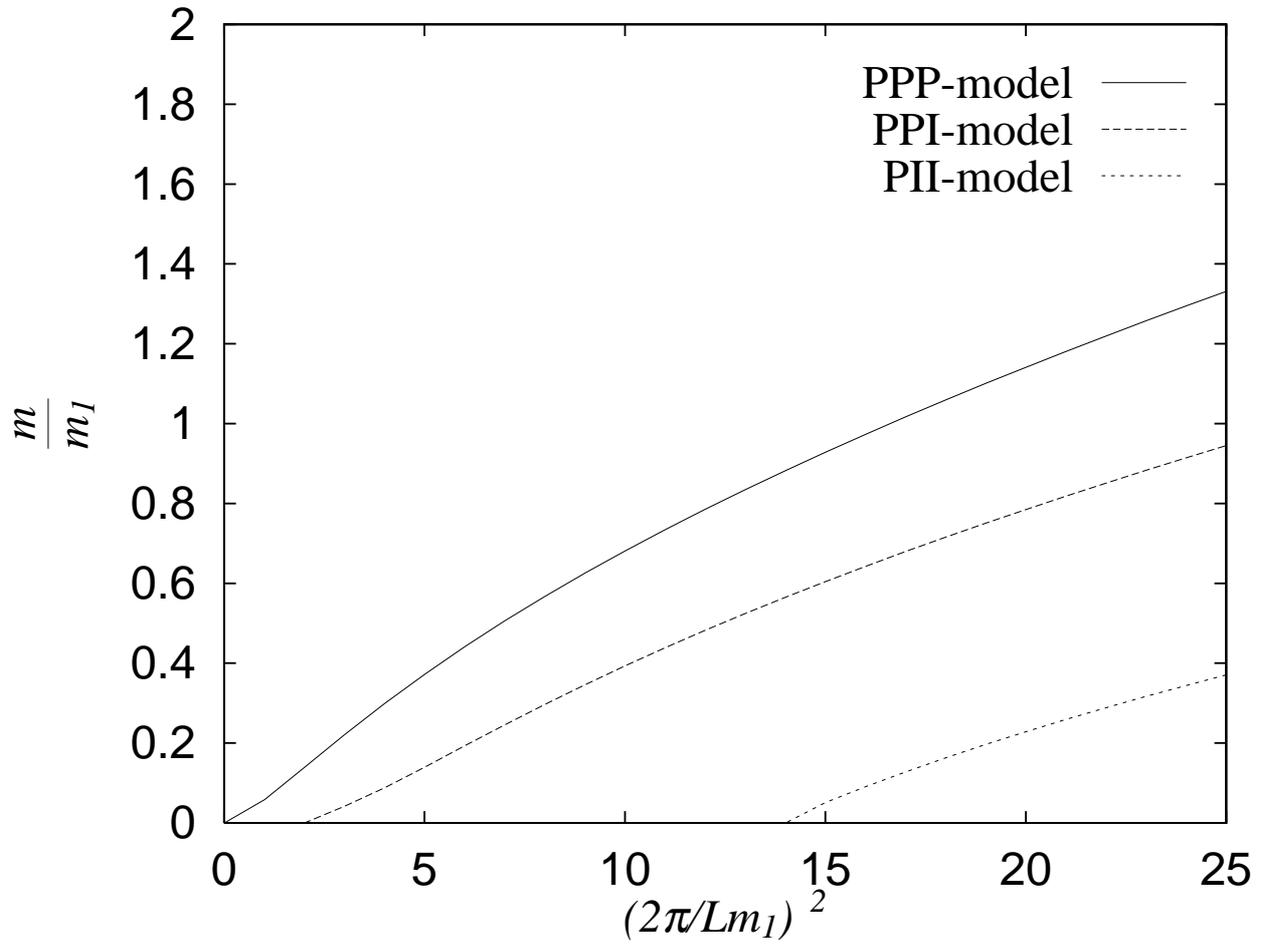
\begin{figure}[H]
\caption[]{fermion mass of the models with $\lambda_{\rm
r}<\lambda_{\rm cr}$. }
\label{fig:figgapPII}
\end{figure}

Fig.\ref{fig:figgapP} shows the case that the coupling constant is larger
than the critical value of Minkowski spacetime, i.e., $\lambda_{\rm r}
>\lambda_{\rm cr}$, and that the phase is broken in the limit 
$L\rightarrow\infty$.
The phase structure of these fields with the 
periodic boundary condition has the interesting feature that 
the phase becomes broken due to the compactified
space even when the coupling constant is smaller than
the critical value and the phase is symmetric in the limit of 
$L\rightarrow\infty$. 
In order to show this, let us introduce a ``mass'' $\mul$ instead of 
the coupling constant by 
\begin{equation}
  \biggl\vert{1\over\lambda_{\rm r}}-{1\over\lambda_{\rm cr}}\biggr\vert
  +{\tr\over(4\pi)^2}\biggl(\ln{\Lambda^2\over\mul^2}\biggr)
  {\mul^2\over\mu^2}=0.
\end{equation}
Then the gap equation reduces to
\begin{equation}
 \ln{\Lambda^2\over\mul^2}
  +{\m^2\over\mul^2}\biggl(\ln{\Lambda^2\over\mul^2}-
  \ln{\m^2\over\mul^2}\biggr)
 -4\sum_{\bn\neq0}{\alpha(\bn)\over(\mul\zetan)}
  {\m\over\mul}K_{1}(\m\zetan)=0~.
\end{equation}
We show the solution of the gap equation 
in Fig.\ref{fig:figgapPII}.
Here the cut-off parameter is chosen as $\Lambda/\m_1=10$.
The fermion mass becomes non-zero value and the broken
phase appears as $L$ becomes smaller in the compactified spaces.
Thus the effect of compactified space makes the phase broken
when the periodic boundary condition is considered.


\subsection{Antiperiodic and periodic boundary conditions}

The above investigation suggests that the dynamical phase of the fields
is significantly affected by the compactification of the spatial sector.
The effect works in different way according to the 
boundary conditions of the fermion fields.
In this subsection, we consider the models in which the different
boundary conditions are adopted for different directions of compact
space, to check these features in more detail.

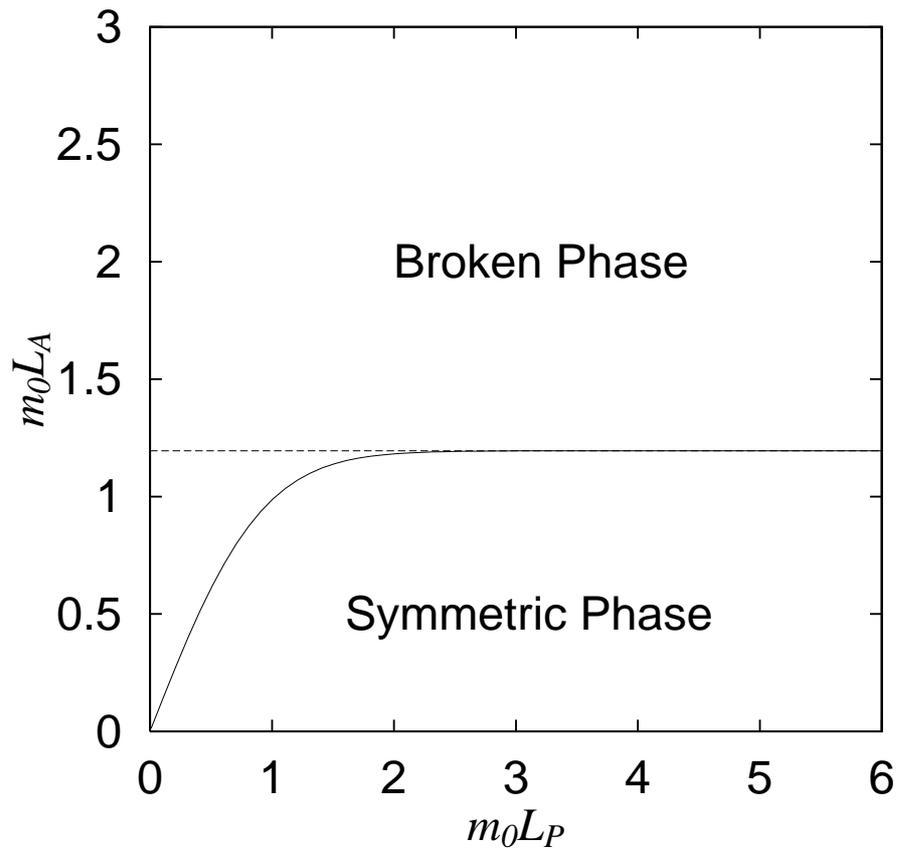
\begin{figure}[H]
\caption[]{The phase diagram for the PAI-model.}
\label{fig:phasePAImodel}
\end{figure}

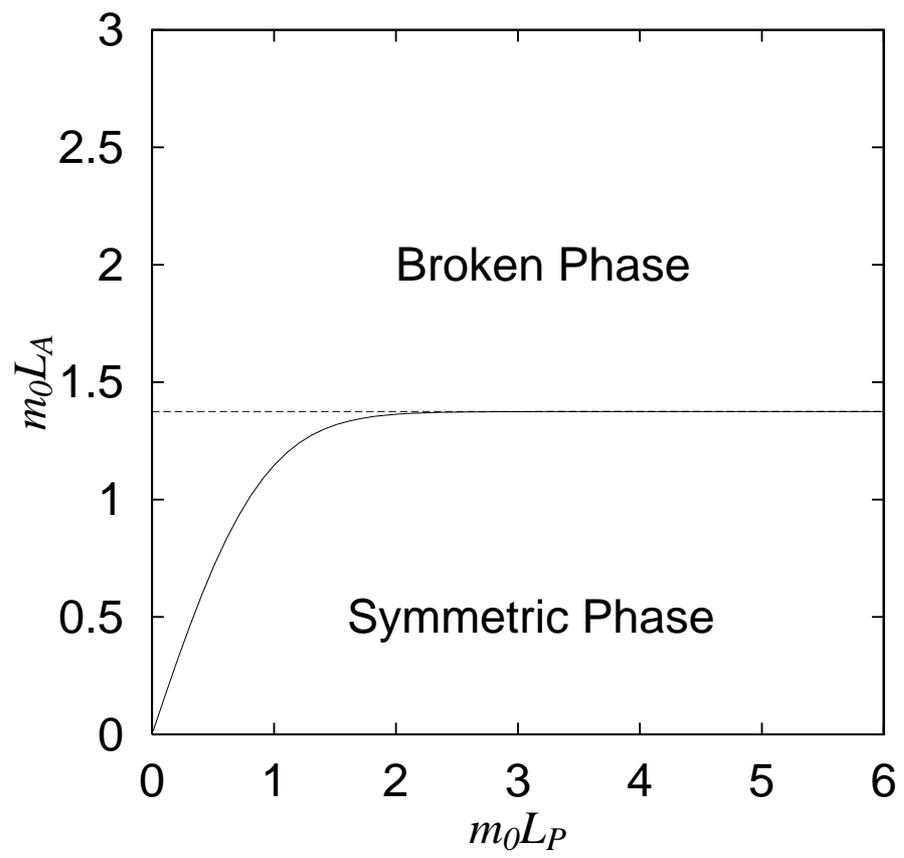
\begin{figure}[H]
\caption[]{The phase diagram for the PAA-model.}
\label{fig:phasePAAmodel}
\end{figure}

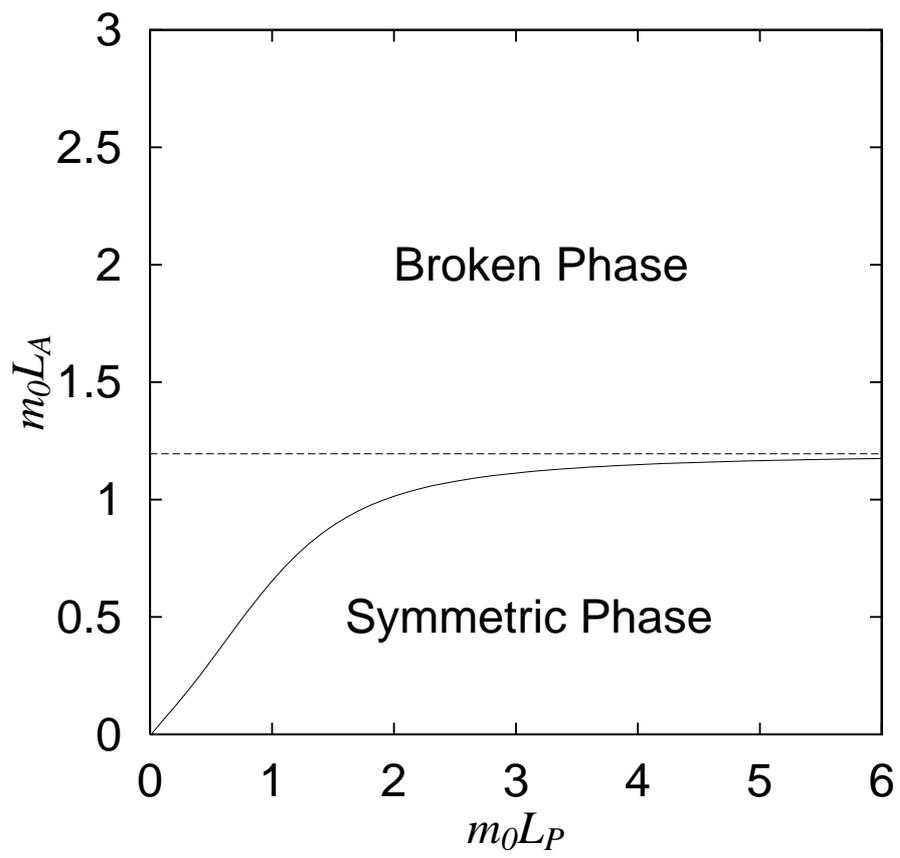
\begin{figure}[H]
\caption[]{The phase diagram for the PPA-model.}
\label{fig:phasePPAmodel}
\end{figure}

First we consider the model that the periodic boundary condition is
imposed in the $x_1$-direction in the period $L_1=\LP$ and
the antiperiodic boundary condition in the $x_2$-direction
in the period $L_2=\LA$ and the other side is infinite, i.e., $L_3=\infty$.
We call this case PAI-model. 
Fig.\ref{fig:phasePAImodel} shows the phase diagram of the PAI-model
,which has the broken phase in the limit of Minkowski
space, in the ($\LP-\LA$) plane. 
The critical value of $\LA$ at large $\LP$ is $m_0L_A\simeq1.20$.
We find  the similar behavior in section 3-B.
We next consider the model that
the periodic boundary condition is imposed in the $x_1$-direction 
in the period $L_1=\LP$ and the antiperiodic boundary condition 
in the $x_2$- and $x_3$-direction in the period $L_2=L_3=\LA$, which 
we call this PAA-model. 
Fig.\ref{fig:phasePAAmodel} is the phase diagram of PAA-model
in the  ($\LP-\LA$) plane. 
The critical value of $\LA$ at large $\LP$ is $m_0L_A\simeq1.38$.
Finally we consider the PPA-model, i.e.,
the periodic boundary condition is imposed in the $x_1$- and $x_2$-direction 
in the period $L_1=L_2=\LP$ and the antiperiodic boundary condition 
in the $x_3$-direction in the period $L_3=\LA$. 
Fig.\ref{fig:phasePPAmodel} is the phase diagram of PPA-model.
The critical value of $\LA$ at large $\LP$ is same as that of PAI-model
All these models show the similar behavior in the subsection 3-B.

%


To end this section, we summarize the results.
We have investigated the nature of the
effective potential in the compactified space in four dimension.
The effect of the compactification of the spatial sector 
changes the phase structure of the four-fermion theory.
The consequent results on the effective potential comes up in
different way according to the boundary condition of the
fermion fields. The antiperiodic boundary condition 
tends to restore the symmetry and the periodic boundary condition
does to break the symmetry, when the effect of the compact space
becomes large. Thus we can construct the both models that 
the symmetry is broken and restored in the course of the 
expansion of the universe.

\section{Summary and Discussions}

We have investigated the four-fermion theory in compact 
flat space with non-trivial topology.
By using the effective potential and the gap equation 
in the leading order of the $1/N$ expansion we find
the phase structure of the theory in three and four
spacetime dimensions.
In three dimensions three class of models are considered
according to the variety of the boundary conditions
for fermion fields. The phase structure
of the theory is examined for three models
When taking  the antiperiodic boundary condition,
the broken chiral symmetry tends to be restored for a sufficiently
small $L$. The phase transition is of the second order.
When taking the periodic boundary condition, the chiral 
symmetry tends to be broken down for a small $L$.
In four dimensions we also see the same effects appears
in the compact flat spaces.

Therefore the drastic change of the phase structure 
is induced by the compact space with no curvature.
In the torus space with the antiperiodic boundary condition,
the finite size effect decreases the dynamical fermion mass and
the chiral symmetry is restored for a small universe 
beyond some critical size ($L < L_{cr}$).
On the other hand, the torus space with the 
periodic boundary condition for fermion fields,
the finite size effect causes the 
opposite influence to the phase structure.
The dynamical fermion mass is increased as the size 
$L$ decreases and the chiral symmetry may be broken
down for $L < L_{cr}$, even when we set
symmetric phase at $L\rightarrow\infty$.
In some cases only the broken phase is observed
for any finite $L$ ($L_{cr}\rightarrow \infty$)
even if the coupling constant $\lambda_{r}$ of the
four-fermion interactions is sufficiently small.

According to the behavior of the effective potential in section 3, 
the value of the vacuum energy in the true vacuum becomes lower
when the space size (volume) of the compactified direction becomes small
in the model associated with the periodic boundary condition.
On the contrary, for the model associated with the 
the antiperiodic boundary condition,
the value of the the vacuum energy in the true vacuum 
is raised, when the space size (volume) becomes small.
Therefore we find that the effect of the periodic 
boundary condition forces attractively (negative pressure)
and that of antiperiodic boundary
condition forces repulsively (positive pressure).
This fact resembles to the well known 
Casimir effect \cite{CS}, which is found in QED.
The Casimir effect gives rise to the vacuum pressure
due to the effect of the finite volume. 

Using the momentum space representation we understand 
these effects of boundary conditions in the following way.
In the compactified space the momentum is discritized.
The fermion fields with an antiperiodic boundary condition
can not take a momentum smaller than ($ |p| \geq \pi/L$).
Thus the possible momentum of the internal fermion fields
becomes larger when $L$ becomes small.
Since the lower momentum fermion field 
has played an essential role to break the chiral
symmetry, the vacuum expectation value of the 
composite operator $\langle\bar{\psi}\psi\rangle$
disappears for a sufficiently small $L$ and the broken
symmetry is restored.
Contrary to this  
the fermion fields with a periodic boundary condition
can take a vanishing momentum even if the space is compact.
Hence the finite size of $L$ has no effect to restore the symmetry.
In the compactified space
the fermion field $\psi(x)$ can interact with
the field $\psi(x+nL)$ for a compactified direction.
Thus the finite size effect seems to make the interaction stronger.
Summing up all the correlations
$\langle\bar{\psi}(x)\psi(x+nL)\rangle$,
the vacuum expectation value of the composite field
becomes larger as $L$ decreases.
We can understand it as the dimensional reduction.
Compactifying one direction to the size $L$ in $D$-dimensional
space, it looks $D-1$-dimensional space for particles
with Compton wavelength much larger than the size $L$.
In the lower dimensional space the influence from 
the lower momentum fermion exceeds. Then  
the finite size effect 
breaks the chiral symmetry for the model with the 
periodic boundary condition.

From the cosmological point of view, 
some mechanism, e.g., inflation, is needed to explain the hot and 
large universe.
Even in the torus universe, inflation seems to be needed
to solve the horizon problem.
As had been discussed in Ref.\cite{Branden},
we need some special idea to lead inflation in the
context of the dynamical symmetry breaking scenario.
An inflation model which is induced 
by a composite fermion field in the
context of the dynamical symmetry breaking scenario
is investigated in some class of supersymmetric particle model \cite{preon}.
We will need further investigation to test the 
symmetry breaking in the early universe.

\subsection*{ACKNOWLEDGMENTS}
We would like to thank professor T.~Muta for
useful discussions and  comments.
We also thanks M.~Siino for helpful comments.

\appendix
\section*{}

In the present paper we have calculated the effective potential
starting from the Feynman propagator defined by 
Eq.(\ref{defGF}).
The spacetime considered here reduces to the cylindrical 
universe $R\otimes S^{1}$ for $D=2$.
The effective potential of the four-fermion theory 
in a cylindrical universe can be obtained by using another 
representation of the Feynman propagator.\cite{KNSY,IIM}
To justify our method
we calculate the effective potential (\ref{def:V:topo})
in two dimensions and compare it with the result
given in Ref.\cite{IIM}.

Taking the two dimensional limit Eqs.(\ref{Vivren}) 
and (\ref{exVfv}) read
\begin{equation}
     \VivR =\frac{1}{2}\left(\frac{1}{\lambda_{r}}
           -\frac{3\tr}{4\pi}\right)\sigma^{2}
           +\frac{\tr}{8\pi}\sigma^{2}
           \ln\left(\frac{\sigma}{\mu}\right)^{2}\, ,
\label{epot:Min:2D}
\end{equation}
\begin{equation}
     \Vfv =\sum^{\infty}_{n=1}\alpha(n)\frac{\tr}{\pi n L}
          |\sigma |\left[
          K_{1}\left(|\sigma | n L\right)
          -\frac{1}{|\sigma | n L}\right]\, ,
\end{equation}
respectively.
Eq.(\ref{epot:Min:2D}) is exactly equal to
the effective potential of Gross-Neveu model
in two-dimensional Minkowski space.
Thus the other part of the effective potential
$\Vfv$ gives the finite size and topological 
effect also in $R\otimes S^{1}$.
To compare the results with the one obtained by another
method we change the expression of $\Vfv$.

The effective potential $\Vfv$ can be written as
\begin{equation}
    \Vfv =-\frac{\tr}{\pi}\int^{\sigma}_{0}s\ ds
    \sum^{\infty}_{n=1}\alpha(n)K_{0}(n|s|L)\, .
\label{D2:Vfv}
\end{equation}
Now we use the formula
\begin{eqnarray}
    I(x)&=&\sum^{\infty}_{n=1}\alpha(n)K_{0}(n x)\, , \nonumber \\
        &=&\sum^{\infty}_{n=1}\alpha(n)
           \int^{\infty}_{0}dt\frac{\cos(nxt)}{\sqrt{t^{2}+1}}\, ,\\
\label{app:formula}
        &=&\int^{\infty}_{0}dt\frac{1}{2\sqrt{t^{2}+1}}
           \left(\sum^{\infty}_{n=-\infty}
           e^{in\beta(xt)}-1\right)\, ,\nonumber
\end{eqnarray}
where the variable $\beta(xt)$ depends on the boundary condition
and given by
\begin{equation}
     \beta(xt)=\left\{
     \begin{array}{ll}
          xt\, ,     & \mbox{periodic boundary condition}\, ,     \\
          xt+\pi\, , & \mbox{antiperiodic boundary condition}\, .
     \end{array}
     \right.
\label{beta:app}
\end{equation}
Performing the summation in Eq.($A4$), the function $I(x)$
reads
\begin{equation}
     I(x)=\int^{\infty}_{0}d\omega\frac{1}{\sqrt{\omega^{2}+x^{2}}}
          \left(\pi\delta_{p}(\beta(\omega))-\frac{1}{2}\right),
\label{app:i:2}
\end{equation}
where $\delta_{p}(x)$ is the periodic delta function
defined by
\begin{equation}
     \delta_{p}(x)=\frac{1}{2\pi}
     \sum^{\infty}_{n=-\infty}e^{i n x}\, .
\end{equation}
To obtain (\ref{app:i:2}) we change the integration variable $t$
to $\omega =xt$.
According to the property of $\delta_{p}(x)$
Eq.(\ref{app:i:2}) can be represented in the following form
\begin{equation}
     I(x)=\sum^{\infty}_{n=0}\frac{\pi}{\sqrt{\omega_{n}^{2}+x^{2}}}
     -\frac{1}{2}\int^{\infty}_{0}
     d\omega\frac{1}{\sqrt{\omega^{2}+x^{2}}}\, ,
\label{app:i:3}
\end{equation}
where $\omega_{n}$ is given by
\begin{equation}
     \omega_{n}=\left\{
     \begin{array}{ll}
          2n \pi\, ,     & \mbox{periodic boundary condition}\, ,     \\
          (2n+1) \pi\, , & \mbox{antiperiodic boundary condition}\, .
     \end{array}
     \right.
\label{omega:app}
\end{equation}
Substituting Eq.(\ref{app:i:3}) to Eq.(\ref{D2:Vfv}),
the integrand of the Eq.(\ref{D2:Vfv}) is modified as
\onecol
\begin{equation}
     \Vfv =-\frac{\tr}{\pi}\int^{\sigma}_{0}s\ ds\left[\frac{\pi}{L}
     \sum^{\infty}_{n=0}\frac{1}{\sqrt{(\omega_{n}/L)^{2}+s^{2}}}
     -\frac{1}{2}\int^{\infty}_{0}
     d\omega\frac{1}{\sqrt{(\omega/L)^{2}+s^{2}}}
     \right]\, .
\label{2d:Vfv:2}
\end{equation}
We can easily perform the integration over $s$ and find
\begin{equation}
     \Vfv =-\frac{\tr}{L}
     \sum^{\infty}_{n=0}\left(\sqrt{\left(
     \frac{\omega_{n}}{L}\right)^{2}+\sigma^{2}}
     - \left|\frac{\omega_{n}}{L}\right|\right)
     +\frac{\tr}{2\pi}\int^{\infty}_{0}
     d\omega\left(\sqrt{\left(\frac{\omega}{L}\right)^{2}+\sigma^{2}}
     -\left|\frac{\omega}{L}\right|\right)\, .
\label{2d:Vfv:3}
\end{equation}
\twocol

Using the momentum space representation of the Feynman 
propagator the effective potential of the Gross-Neveu model 
is given by
\begin{equation}
     \Viv =\frac{1}{2\lambda_{0}}\sigma^{2}
     -\tr\int^{\infty}_{0}\frac{dk}{2\pi}
     \left(\sqrt{k^{2}+\sigma^{2}}
     -\left|k\right|\right)\, ,
\label{2d:GN:app}
\end{equation}
in two-dimensional Minkowski spacetime.
Inserting Eqs.(\ref{2d:Vfv:3}) and (\ref{2d:GN:app})
into Eq.(\ref{def:V:topo}) the effective potential 
in $R\otimes S^{1}$ reads
\begin{equation}
     V =\frac{1}{2\lambda_{0}}\sigma^{2}
     -\frac{\tr}{L}
     \sum^{\infty}_{n=0}\left(\sqrt{\left(
     \frac{\omega_{n}}{L}\right)^{2}+\sigma^{2}}
     - \left|\frac{\omega_{n}}{L}\right|\right)\, .
\label{v:2d:fine}
\end{equation}

In the case of the antiperiodic boundary condition
Eq.(\ref{v:2d:fine}) reproduces the result obtained 
in Ref.\cite{KNSY,IIM}.
It is well-known that the field theory in $R\otimes S^{1}$
is equivalent to the finite-temperature field theory
\footnote{In the finite temperature field theory
it is forbidden to choose the periodic boundary
condition for a fermion field}.
As is shown in Ref.\cite{IIM}, the effective potential 
(\ref{v:2d:fine}) is in agreement with that of the
finite temperature four-fermion theory in $D=2$
with recourse to the relation between the size of the universe
$L$ and the temperature $T$
\begin{equation}
    L=\frac{1}{k_{B}T}
\end{equation}
with $k_{B}$ the Boltzmann constant.
In this case it is known that the broken chiral symmetry
is restored for sufficiently small $L < L_{cr}$ at the large $N$
limit. The critical size of the universe is given by
\begin{equation}
     L_{cr}m_{0}=\pi e^{-\gamma}\, ,
\end{equation}
where $\gamma$ is the Euler constant.
It is equal to the well-known formula of the critical
temperature for the Gross-Neveu model at finite 
temperature\cite{FT}.

On the other hand the symmetric phase is not observed
for any values of the coupling constant $\lambda_{r}$
and $L$ in the case of the periodic boundary condition
as is shown below.
Using the Eqs. (A1) and (A11),
the gap equation of the present theory is given by
\onecol
\begin{equation}
     \frac{dV}{d\sigma}
     =\sigma\left[
     \frac{1}{\lambda_{r}}-\frac{\tr}{4\pi}\left(2
     -\ln\left|\frac{\sigma}{\mu}\right|^{2}\right)
     -\sum^{\infty}_{n=0}\frac{\tr}{\sqrt{\omega_{n}^{2}+(L\sigma)^{2}}}
     -\frac{\tr}{2\pi}\int^{\infty}_{0}
     d\omega\frac{1}{\sqrt{(\omega/L)^{2}+\sigma^{2}}}
     \right]=0\, .
\end{equation}
\twocol
If we take the limit, $\sigma\rightarrow +0$, we find that
\begin{equation}
     \left.\frac{dV}{d\sigma}\right|_{\sigma\rightarrow +0}
     \rightarrow -\frac{1}{L}\, .
\end{equation}
The derivative of the effective potential has
a negative value at $\sigma\rightarrow +0$.
Thus the minimum of the effective potential is located at
non-vanishing $\sigma$ and the chiral symmetry is always
broken down for the fermion field with periodic boundary
condition irrespective of the value of the coupling 
constant $\lambda_{r}$.
Evaluating the effective potential numerically we see that
the dynamically generated fermion mass becomes heavier
as the size $L$ decreased.
Therefore the boundary condition change the finite size 
effect conversely also in two dimensions and our method
agrees with the well-known results. 


\onecol

\newpage

\pagestyle{empty}
\begin{center}
{\bf Figure Captions}
\end{center}

\begin{figure}[H]
\caption[]{Solution of the gap equation for the AA-model.}
\label{fig:Gap3dAA}
\end{figure}

\begin{figure}[H]
\caption[]{Behavior of the effective potential for the AA-model
in the case of $L_{1}=L_{2}=L$. 
We have set $\lambda_{r}=2 \lambda_{cr}$, and used the 
notation  $\hat{k}\equiv k/\mu^{2}$,
$\hat{\sigma}\equiv \sigma/\mu$,
$\hat{V}\equiv V/(\tr\mu^{3})$.
The critical value is $k_{cr}/m_{0}^{2} \approx 15.1$. }
\label{fig:EffptAA}
\end{figure}

\begin{figure}[H]
\caption[]{The phase diagram for the AA-model. }
\label{fig:PhaseAALxLy}
\end{figure}

\begin{figure}[H]
\caption[]{Solution of the gap equation for the PA-model.}
\label{fig:Gap3dPA}
\end{figure}

\begin{figure}[H]
\caption[]{Behavior of the effective potential for the PA-model
in the case of $L_{1}=L_{2}=L$.
Here we have set  $\lambda_{r}=2 \lambda_{cr}$,
 $\hat{k}\equiv k/\mu^{2}$,
 $\hat{\sigma}\equiv \sigma/\mu$,
 $\hat{V}\equiv V/(\tr\mu^{3})$.
 The critical value is $k_{cr}/m_{0}^{2} \approx 30.3$. }
\label{fig:EffptPA}
\end{figure}

\begin{figure}[H]
\caption[]{Behavior of the effective potential for the PA-model
with $L_1$ fixed by $\hat{k}_{1}=10.0 $.
Here we have set  $\lambda_{r}=2 \lambda_{cr}$,
 $\hat{k}_{i}\equiv k_{i}/\mu^{2}$,
 $\hat{\sigma}\equiv \sigma/\mu$,
 $\hat{V}\equiv V/(\tr\mu^{3})$.
 The critical value is $k_{2 \,cr}/m_{0}^{2} \approx 21.9$. }
\label{fig:EffptPAKx10KyetcAP}
\end{figure}

\begin{figure}[H]
\caption[]{Behavior of the effective potential for the PA-model
with $L_2$ fixed by $\hat{k}_{2}=10.0$.
Here we have set  $\lambda_{r}=2 \lambda_{cr}$,
 $\hat{k}_{i}\equiv k_{i}/\mu^{2}$,
 $\hat{\sigma}\equiv \sigma/\mu$,
 $\hat{V}\equiv V/(\tr\mu^{3})$.}
\label{fig:EffptPAKxetcKy10AP}
\end{figure}

\begin{figure}[H]
\caption[]{The phase diagram for the PA-model.}
\label{fig:PhasePALxLy}
\end{figure}

\begin{figure}[H]
\caption[]{Behavior of the gap equation for the PP-model.}
\label{fig:Gap3dPP}
\end{figure}

\begin{figure}[H]
\caption[]{Behavior of the effective potential for the PP-model
in the case of $L_{1}=L_{2}=L$. The notations are same as 
Fig.(\ref{fig:EffptPA}).
}
\label{fig:EffptPP}
\end{figure}

\begin{figure}[H]
\caption[]{Solution of the gap equation for 
the models with antiperiodic boundary condition.
}
\label{fig:figgapA}
\end{figure}

\begin{figure}[H]
\caption[]{The cut-off dependence of the critical value  of
$(2\pi/L_{\rm cr}m_0)^2$.}
\label{fig:figcrtA}
\end{figure}

\begin{figure}[H]
\caption[]{Solution of the gap equation for 
the models with periodic boundary condition.
}
\label{fig:figgapP}
\end{figure}

\twocol

\onecol

\end{document}